\documentclass[twocolumn,english]{revtex4}
\usepackage[T1]{fontenc}
\usepackage[utf8]{luainputenc}
\usepackage{amsmath}
\usepackage{amssymb}
\usepackage{graphicx}
\usepackage{esint}

\makeatletter
\@ifundefined{textcolor}{}
{%
 \definecolor{BLACK}{gray}{0}
 \definecolor{WHITE}{gray}{1}
 \definecolor{RED}{rgb}{1,0,0}
 \definecolor{GREEN}{rgb}{0,1,0}
 \definecolor{BLUE}{rgb}{0,0,1}
 \definecolor{CYAN}{cmyk}{1,0,0,0}
 \definecolor{MAGENTA}{cmyk}{0,1,0,0}
 \definecolor{YELLOW}{cmyk}{0,0,1,0}
}

\usepackage{babel}

\usepackage{babel}

\usepackage{babel}

\usepackage{babel}

\usepackage{babel}

\usepackage{babel}

\usepackage{babel}

\usepackage{babel}

\usepackage{babel}

\usepackage{babel}

\usepackage{babel}

\usepackage{babel}

\usepackage{babel}

\makeatother

\usepackage{babel}
\begin{document}

\title{Quench between a Mott insulator and a Lieb-Liniger liquid}

\author{Garry Goldstein and Natan Andrei}

\address{Department of Physics, Rutgers University}

\address{Piscataway, New Jersey 08854}
\begin{abstract}
In this work we study a quench between a Mott insulator and a repulsive
Lieb-Liniger liquid. We find explicitly the stationary state when
a long time has passed after the quench. It is given by a GGE density
matrix which we completely characterize, calculating the quasiparticle
density describing the system after the quench. In the long time limit
we find an explicit form for the local three body density density
density correlation function and the asymptotic long distance limit
of the density density correlation function. The latter is shown to
have gaussian decay at large distances. 
\end{abstract}
\maketitle

\section{\label{sec:Introduction}Introduction}

The question of whether an isolated quantum system equilibrates is
of fundamental interest to our understanding of nonequilibrium dynamics.
This question is extremely difficult to study in the condensed matter
context as there are few truly isolated systems. However recent advances
in the study of ultracold atoms have brought this question to the
forefront. Thanks to an unprecedented amount of tunability and isolation
in cold atom systems it has become possibly to study complex nonequilibrium
dynamics of many body systems. Triggered by spectacular experimental
advances \cite{key-1,key-4,key-5,key-6,key-7,key-8,key-9,key-3,key-11}
there has been great effort recently in the theoretical study of thermalization
of isolated systems \cite{key-12,key-13,key-14,key-15,key-16,key-17,key-18,key-19,key-8-1}.
The most important questions asked are whether a system equilibrates,
what are the dynamics of the equilibration process and what ensemble
if any describes the long time limit of an isolated system?

One of the most surprising recent experimental and theoretical results
is that at long times the quenched Lieb-Liniger gas \cite{key-9-1}
retains memory of its initial state \cite{key-10-1,key-8-1,key-5}
and does not appear to relax to thermodynamic equilibrium. This is
due to the fact that the Lieb Liniger hamiltonian: 
\begin{equation}
H_{LL}=\intop_{-\infty}^{\infty}dx\left\{ \partial_{x}b^{\dagger}\left(x\right)\partial_{x}b\left(x\right)+c\left(b^{\dagger}\left(x\right)b\left(x\right)\right)^{2}\right\} ,\label{eq:lieb-lin-hamiltonian}
\end{equation}
has an infinite number of conserved charges $I_{i}$. Here $b^{\dagger}\left(x\right)$
is the bosonic creation operator at the point $x$ and $c$ is the
coupling constant. These conserved quantities in turn imply that there
is a complete system of eigenstates for the Lieb Liniger gas which
may be parametrized by sets of rapidities $\left\{ k_{i}\right\} $.
To understand the equilibration of this gas it was recently proposed
that it is insufficient to consider only thermal ensembles but it
is also necessary to include these nontrivial conserved quantities.
It was shown \cite{key-10-1,key-8-1} that the gas relaxes to a state
given by the generalized Gibbs ensemble GGE with its density matrix
being given by 
\begin{equation}
\rho_{GGE}=\frac{1}{Z}\exp\left(-\sum\alpha_{i}I_{i}\right)\label{eq:GGE_density_matrix-1}
\end{equation}

Where the $I_{i}$ are the conserved quantities given by $I_{i}\left|\left\{ k\right\} \right\rangle =\sum k^{i}\left|\left\{ k\right\} \right\rangle $
and the $\alpha_{i}$ are the generalized inverse temperatures and
$Z$ is a normalization constant insuring $Tr\left[\rho_{GGE}\right]=1$.
It was shown that correlation functions of the Lieb-Liniger gas at
long times may be computed by taking their expectation value with
respect to the GGE density matrix, e.g. $\left\langle \Theta\left(t\rightarrow\infty\right)\right\rangle =Tr\left[\rho_{GGE}\Theta\right]$.
It was also later shown \cite{key-1-1} that the the GGE ensemble
is equivalent to a pure state $\rho_{GGE}\cong\left|\vec{k}_{0}\right\rangle \left\langle \vec{k}_{0}\right|$
for an appropriately chosen $\left|\vec{k}_{0}\right\rangle $.

\begin{figure}
\begin{centering}
\includegraphics[width=1\columnwidth]{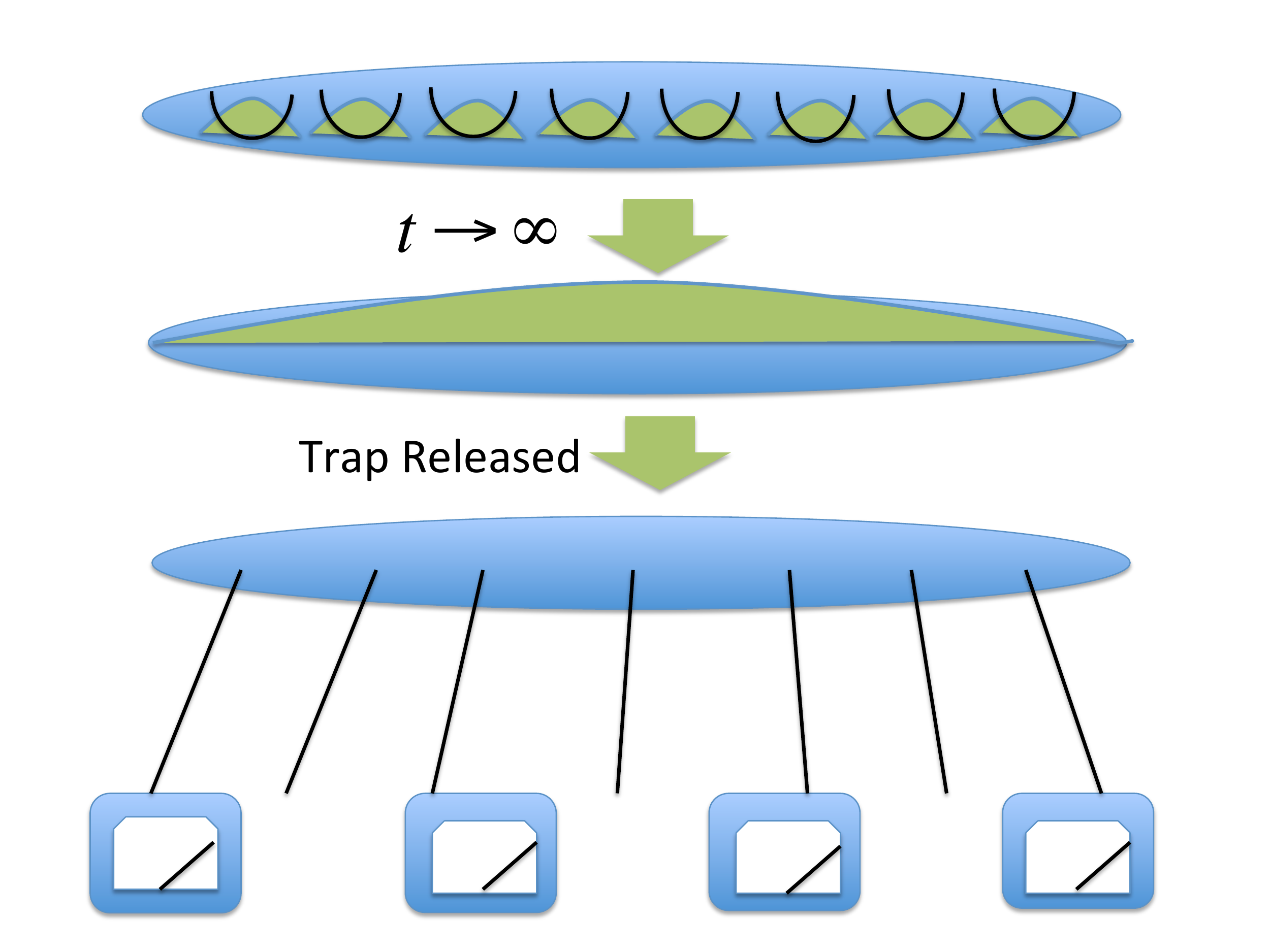} 
\par\end{centering}

\protect\protect\protect\protect\caption{\label{fig:Trap_release}The Lieb Liniger gas is initialized in a
nonequilibrium state - the Mott insulator (say by applying an external
lattice). The lattice is removed and the system is allowed to relax
for a long time. Its correlation functions are then measured.}
\end{figure}

In this work we consider the repulsive Lieb-Liniger model. We consider
the case when the system is initialized in a deep lattice. So the
state of the system is well described by a Mott insulator: 
\begin{equation}
\left|\Phi_{0}\right\rangle =\prod_{j=-\infty}^{\infty}\int_{-\infty}^{\infty}\varphi\left(x+jl\right)b^{\dagger}\left(x\right)\left|0\right\rangle \label{eq:Initial_state}
\end{equation}
with $\varphi\left(x\right)=\frac{e^{-x^{2}/\sigma}}{\left(\pi\sigma/2\right)^{1/4}}$,
with $\sqrt{\sigma}\ll l$. The lattice is then released so that the
system evolves under the Hamiltonian given in Eq. (\ref{eq:lieb-lin-hamiltonian}).
The Lieb-Liniger gas is allowed to relax for a long time, $\left|\Psi\right\rangle =e^{iH_{LL}t}\left|\Phi_{0}\right\rangle $,
thereby establishing a GGE with $\rho_{GGE}\cong\left|\Psi\right\rangle \left\langle \Psi\right|$,
see figure (\ref{fig:Trap_release}). We note that using a saddle
point approach the authors in \cite{key-1-1} were able to identify
the limit value $\left|\Psi\right\rangle \left\langle \Psi\right|$
with $\left|\vec{k}_{0}\right\rangle \left\langle \vec{k}_{0}\right|$,
where the quasiparticle distribution characterizing $\vec{k}_{0}$
satisfies the appropriate Bethe-Ansatz equation (see below).

In this work we establish the exact quasiparticle density of this
GGE. We use it to calculate various correlation functions for the
final state. In particular we find a Gaussian decay of the density
density correlation function at large distances for the final state.
Overall we obtain a complete characterization of the final state.
Our results are applicable for any coupling constant $c>0$. We would
like to note that a quench between a BEC and a Lieb-Liniger gas has
been previously considered in \cite{key-3-1}. The rest of the paper
is organized as follows. In Section \ref{sec:Quasiparticle_density}
we calculate the quasiparticle density of the final state; in Section
\ref{sec:Local-correlation-functions} we compute the two and three
body local correlation functions for the final state; in Section \ref{sec:Density-Density-correlations}
we compute the long distance density density correlation functions
of the final state, here we also generalize our results (we see that
the final correlation functions are heavily independent of certain
types of disorder) and in Section \ref{sec:Conclusions} we conclude.

\section{\label{sec:Quasiparticle_density}Quasiparticle density}

We would like to characterize the final state obtained in the quench
between a Mott insulator and the Lieb-Liniger gas, it is given by
a GGE. For the Lieb-Liniger gas it is possible to reduce the GGE density
matrix to a pure state \cite{key-1-1}. That is for any local correlation
function $\Theta$: 
\begin{equation}
tr\left[\Theta\rho_{GGE}\right]=\left\langle \vec{k}_{0}\right|\Theta\left|\vec{k}_{0}\right\rangle \label{eq:GGE_State_equivalence}
\end{equation}
for an appropriately chosen eigenstate of the Lieb Liniger Hamiltonian
$\left|\vec{k}_{0}\right\rangle $. Furthermore following the authors
of \cite{key-1-1} we are able to specify this pure state $\left|\vec{k}_{0}\right\rangle $.
To do so let us denote by $L\rho_{p}\left(k\right)dk$ as the number
of particles in the interval $\left[k,k+dk\right]$, $L\rho_{h}\left(k\right)dk$
as the number of holes in the interval $\left[k,k+dk\right]$ and
$L\rho_{t}\left(k\right)dk$ as the number of states in the interval
$\left[k,k+dk\right]$ so that $\rho_{t}\left(k\right)=\rho_{p}\left(k\right)+\rho_{h}\left(k\right)$.
Here $L$ is the length of the system. Then the density of particles
in k-space corresponding to $\left|\vec{k}_{0}\right\rangle $ is
given by the following equations \cite{key-1-1}: 
\begin{equation}
L\int dk\rho_{p}\left(k\right)k^{n}=I_{n}\left(t=0\right).\label{eq:Equation of state}
\end{equation}

Therefore to characterize the density of quasiparticles of the final
sate one needs to compute the various conserved quantities $I_{n}\left(t=0\right)$.
To do so note that the conserved charges $I_{n}$ are given by integrals
of local densities: 
\begin{equation}
I_{n}=\int dxJ_{n}\left(x\right)\label{eq:Local_quantites}
\end{equation}
with the local densities being given by $J_{0}\left(x\right)=b^{\dagger}\left(x\right)b\left(x\right)$,
$J_{1}\left(x\right)=ib^{\dagger}\left(x\right)\partial_{x}b\left(x\right)$,
$J_{2}\left(x\right)=\partial_{x}b^{\dagger}\left(x\right)\partial_{x}b\left(x\right)+c\left(b^{\dagger}\left(x\right)b\left(x\right)\right)^{2}$,
$J_{3}\left(x\right)=b^{\dagger}\left(x\right)\partial_{x^{3}}^{3}b\left(x\right)-\frac{3c}{2}\left(b^{\dagger}\left(x\right)\right)^{2}\partial_{x}\left(b\left(x\right)\right)^{2}$
etc. Since these changes are local and the wave function in Equation
(\ref{eq:Initial_state}) is given by a sum of local non-overlapping
single particle terms we have that: 
\begin{equation}
I_{n}\left(t=0\right)=\frac{L}{l}\left\langle 0\right|\int dx\varphi\left(x\right)b\left(x\right)I_{n}\int dy\varphi\left(y\right)b^{\dagger}\left(y\right)\left|0\right\rangle \label{eq:conserved_quantity}
\end{equation}

However for a single particle state it is straightforward to calculate
the value of a local conserved quantity; indeed $I_{n}\left|k\right\rangle =k^{n}\left|k\right\rangle $.
From this we obtain that the conserved charges are zero for odd $n$
and for even $n$ they are given by: 
\begin{equation}
I_{n}\left(t=0\right)=\frac{L}{l}\,\left(\frac{2}{\sigma}\right)^{\frac{n}{2}}\,\frac{n!}{2^{\frac{n}{2}}\left(\frac{n}{2}!\right)}\label{eq:Concerved charges}
\end{equation}

In particular these are finite \cite{key-3-1}. Therefore according
to Equation (\ref{eq:Equation of state}) the quasiparticle density
$\rho_{p}\left(k\right)$ is a distribution whose moments are given
by Equation (\ref{eq:Concerved charges}). However it is straightforward
to obtain such a distribution it is given by: 
\begin{equation}
\rho_{p}\left(k\right)=\frac{\sigma^{\frac{1}{2}}}{\pi^{\frac{1}{2}}l}\exp\left(-\frac{k^{2}\sigma}{2}\right)\label{eq:Quasiparticle_density}
\end{equation}

We note that this expression is completely independent of the coupling
constant $c$ and works for all $c>0$. This distribution completely
characterizes all the properties of the final state and completely
determines $\rho_{GGE}$. For future use we would like to compute
the total density of available states $\rho_{t}\left(k\right)$ as
well as the ratio $\frac{\rho_{p}\left(k\right)}{\rho_{t}\left(k\right)}$.
The total quasiparticle density is given by the equation \cite{key-31}:
\begin{equation}
\rho_{t}\left(k\right)=\frac{1}{2\pi}+\frac{1}{2\pi}\int dqK\left(k,q\right)\rho_{p}\left(q\right),\label{eq:Density_available_states}
\end{equation}
here $K\left(k,q\right)=\frac{2c}{c^{2}+\left(k-q\right)^{2}}$. We
note that for $l\gg\sqrt{\sigma}$ $\rho_{t}\left(k\right)\cong\frac{1}{2\pi}$.
Furthermore the occupation probability is given by $f\left(k\right)\equiv\frac{\rho_{p}\left(k\right)}{\rho_{t}\left(k\right)}\cong\frac{2\sqrt{\pi\sigma}}{l}\exp\left(-\frac{k^{2}\sigma}{2}\right)$.
We note that the charges $I_{n}\left(t=0\right)$ grow very rapidly
so are high order charges are important as such the results presented
in \cite{key-20-1} about exponential decay of density density correlations
will not apply as we will see in Eq. (\ref{eq:density_density_correlator}).

\section{\label{sec:Local-correlation-functions}Local correlation functions}

We would like to calculate the local correlation functions $\langle b^{\dagger}\left(0\right)b\left(0\right)\rangle$,
$\langle b^{\dagger}\left(0\right)b^{\dagger}\left(0\right)b\left(0\right)b\left(0\right)\rangle$
and $\langle b^{\dagger}\left(0\right)b^{\dagger}\left(0\right)b^{\dagger}\left(0\right)b\left(0\right)b\left(0\right)b\left(0\right)\rangle$.
The first of these quantities measures the density of particles in
the system the second measures the effect of interactions while the
third measure the the three particle recombination rate. From conservation
of particle number the density of particles is given in the long time
limit by $\rho=\langle b^{\dagger}\left(0\right)b\left(0\right)\rangle=\frac{1}{l}$.
The two body correlation function is given by \cite{key-20}: 
\begin{equation}
\begin{array}[t]{l}
\left\langle b^{\dagger}\left(0\right)b^{\dagger}\left(0\right)b\left(0\right)b\left(0\right)\right\rangle \cong\\
\cong2\int\frac{dk_{1}}{2\pi}\int\frac{dk_{2}}{2\pi}f\left(k_{1}\right)f\left(k_{2}\right)\frac{\left(k_{2}-k_{1}\right)^{2}}{\left(k_{2}-k_{1}\right)^{2}+c^{2}}+.....
\end{array}\label{eq:two_body-1}
\end{equation}
This integral may be done explicitly: 
\begin{align}
\left\langle b^{\dagger}\left(0\right)b^{\dagger}\left(0\right)b\left(0\right)b\left(0\right)\right\rangle  & =\frac{2}{l^{2}}-\frac{2\sqrt{\pi c^{2}\sigma}}{l^{2}}\times\nonumber \\
 & \times\left[\exp\left(\frac{\sigma c^{2}}{4}\right)Erfc\left(\sqrt{\frac{\sigma c^{2}}{4}}\right)\right]\label{eq:two_body}
\end{align}
The three body integrals may be done similarly \cite{key-20}. They
are given by: 
\begin{equation}
\begin{array}[t]{l}
\left\langle b^{\dagger}\left(0\right)b^{\dagger}\left(0\right)b^{\dagger}\left(0\right)b\left(0\right)b\left(0\right)b\left(0\right)\right\rangle \cong\\
\cong6\int dk_{1}dk_{2}dk_{3}f\left(k_{1}\right)f\left(k_{2}\right)f\left(k_{3}\right)\times\\
\qquad\times\frac{\left(k_{2}-k_{1}\right)^{2}}{\left(k_{2}-k_{1}\right)^{2}+c^{2}}\frac{\left(k_{3}-k_{1}\right)^{2}}{\left(k_{3}-k_{1}\right)^{2}+c^{2}}\frac{\left(k_{3}-k_{2}\right)^{2}}{\left(k_{3}-k_{2}\right)^{2}+c^{2}}+....
\end{array}\label{eq:Three_body}
\end{equation}
rSimplifying we can evaluate this expression in various limits. In
the case that $\sigma c^{2}\ll1$ we obtain that 
\begin{equation}
\left\langle b^{\dagger}\left(0\right)b^{\dagger}\left(0\right)b^{\dagger}\left(0\right)b\left(0\right)b\left(0\right)b\left(0\right)\right\rangle \cong\frac{6}{l^{3}}\label{eq:Three_body_high_c}
\end{equation}
In the case that $\sigma c^{2}\gg1$ we obtain that: 
\begin{equation}
\left\langle b^{\dagger}\left(0\right)b^{\dagger}\left(0\right)b^{\dagger}\left(0\right)b\left(0\right)b\left(0\right)b\left(0\right)\right\rangle \cong\frac{9\times2^{\frac{9}{2}}}{l^{3}c^{6}\sigma^{3}}\label{eq:three_body_supprerssion}
\end{equation}
As such we see a strong suppression of the three body decay rates
for large coupling constants $c$.

\section{\label{sec:Density-Density-correlations}Density Density correlations}

The density density correlation function for generic states has been
previously calculated. We can use this to find the density density
correlation function for our GGE. It is given by \cite{key-21}:

\begin{equation}
\left\langle \rho\left(x\right)\rho\left(0\right)\right\rangle =\rho^{2}+\sum_{k=2}^{\infty}\Gamma_{k}\left(x\right)\label{eq:density_density}
\end{equation}
with the dominant term being given by \cite{key-21}: 
\begin{align}
\Gamma_{2}\left(x\right)= & -\frac{1}{4\pi^{2}}\int_{-\infty}^{\infty}dk_{1}\omega\left(k_{1}\right)f\left(k_{1}\right)\int_{-\infty}^{\infty}dk_{2}\omega\left(k_{2}\right)f\left(k_{2}\right)\times\nonumber \\
 & \times\left(\frac{k_{1}-k_{2}+ic}{k_{1}-k_{2}-ic}\right)\left[\frac{p\left(k_{1},k_{2}\right)}{k_{1}-k_{2}}\right]^{2}\exp\left(xp\left(k_{1},k_{2}\right)\right)\label{eq:Gamma_two}
\end{align}
Here $\omega\left(k\right)=\exp\left(-\frac{1}{2\pi}\int_{-\infty}^{\infty}K\left(k,q\right)f\left(q\right)dq\right)$
and the function $p\left(k_{1},k_{2}\right)$ is given by: 
\begin{equation}
p\left(k_{1},k_{2}\right)=-i\left(k_{1}-k_{2}\right)+\int_{-\infty}^{\infty}dt\, f\left(t\right)\, P\left(t,k_{1},k_{2}\right)\label{eq:P_K1_K2}
\end{equation}
The function $P\left(t,k_{1},k_{2}\right)$ is defined through the
following integral equation: 
\begin{align}
1+2\pi\, P\left(t,k_{1},k_{2}\right) & =\left(\frac{k_{1}-t+ic}{k_{1}-t-ic}\right)\left(\frac{k_{2}-t-ic}{k_{2}-t+ic}\right)\times\nonumber \\
 & \times\exp\left(\int K\left(t,s\right)f\left(s\right)P\left(s,k_{1},k_{2}\right)\right).\label{eq:P_T}
\end{align}
In the case when $l\gg\sqrt{\sigma}$ these equations greatly simplify.
Indeed in that case $\omega\left(k\right)\cong e^{-1}$ and $p\left(k_{1},k_{2}\right)=-i\left(k_{1}-k_{2}\right)$.
With these simplifications the integral in Eq. (\ref{eq:Gamma_two})
becomes, 
\begin{equation}
\left\langle \rho\left(x\right)\rho\left(0\right)\right\rangle \cong\rho^{2}+\frac{1}{4\pi^{2}e^{2}l^{2}}\exp\left(-\frac{x^{2}}{\sigma}\right)\label{eq:density_density_correlator}
\end{equation}
We see a gaussian decay of correlation functions at large distances.

We would like to note that the final result e.g. the final quasiparticle
density see Eq. (\ref{eq:Quasiparticle_density}) is independent of
positional disorder in the Mott insulator. Indeed if the centers of
the particles in the Mott insulator do not form a uniform lattice
but instead have random locations (which would happen for example
for a partial filling of the first Mott band) none of the conserved
quantities, see Equation (\ref{eq:Concerved charges}), change. In
particular if the particles have an average interparticle separation
$\bar{l}$ then the final state has exactly the same correlations
as those given in the main text except with $l\rightarrow\bar{l}$
see Equations (\ref{eq:two_body}), (\ref{eq:Three_body_high_c}),
(\ref{eq:three_body_supprerssion}) and (\ref{eq:density_density_correlator}).
As such we can characterize these states as well.

\section{\label{sec:Conclusions}Conclusions}

We have studied a quench between a Mott insulator and a Lieb-Liniger
liquid. We have found that at long time the system equilibrates to
a GGE. We have characterized the GGE quasiparticle density exactly
and found it to be gaussian. We were able to compute local two and
three body correlation functions as well as density density correlations.
We have found the density density correlation to decay in a Gaussian
manner with distance. We argued our results to be robust to certain
kinds of disorder. The methods used in this paper open the possibility
to compute the final state after a quench when the initial state is
given by a collection of isolated few body states (such as a collection
of dimer molecules on a lattice). The conserved quantities can be
calculated in that case as well allowing us to obtain the final quasiparticle
density as well as various correlation functions.

The results presented here have direct experimental applicability.
Indeed it is not too difficult to prepare a gas of bosons in the Mott
insulator state \cite{key-21-1}. The three body density density correlations
may be measured through trap loss rates \cite{key-20} or through
the third moment of the number of particles \cite{key-22}. The density
density correlators studied may be measured through time of flight
interferometry \cite{key-21-1}, all directly experimentally relevant.

\textbf{Acknowledgments}: This research was supported by NSF grant
DMR 1006684 and Rutgers CMT fellowship.

\end{document}